# Delta Modeling for Software Architectures

Arne Haber[1], Holger Rendel[1], Bernhard Rumpe[1], Ina Schaefer[2]

[1]Software Engineering
RWTH Aachen University, Germany
http://www.se-rwth.de/

[2]Software Systems Engineering Institut
Technische Universität Braunschweig, Germany
http://www.tu-braunschweig.de/sse

**Abstract:** Architectural modeling is an integral part of modern software development. In particular, diverse systems benefit from precise architectural models since similar components can often be reused between different system variants. However, during all phases of diverse system development, system variability has to be considered and modeled by appropriate means. Delta modeling is a language-independent approach for modeling system variability. A set of diverse systems is represented by a core system and a set of deltas specifying modifications to the core system. In this paper, we give a first sketch of how to apply delta modeling in MontiArc, an existing architecture description language, in order to obtain an integrated modeling language for architectural variability. The developed language, $\Delta$-MontiArc, allows the modular modeling of variable software architectures and supports proactive as well as extractive product line development.

## 1 Introduction

Today, modeling of software architectures is an integral part of the software development process [MT10]. Dividing a system into small manageable parts provides an overview of the system structure, subcomponent dependencies, and communication paths. Architectural modeling allows reasoning about structural system properties in early development stages. Reusing well-defined modular components reduces development costs and increases product quality. Especially embedded system architectures have to be carefully designed, because resources like bandwidth or memory are restricted, product quantities are high, and faults that are detected late are extremely expensive.

Diversity is prevalent in modern software systems, in particular in the embedded systems domain. Systems exist in many different variants simultaneously in order to adapt to their application context or customer needs. Software product line engineering [PBvdL05] is a commercially successful approach to develop a set of systems with well-defined commonality and variability. Product line engineering benefits from architectural modeling, since common components can be reused in different system variants. However, in order to develop a software product line, system variability has to be considered in all phases of the development process, including architectural modeling. This means that the variability of the architectures in different system variants has to be modeled by appropriate means.

*Delta modeling* [CHS10, SBB$^+$10, Sch10] is a transformational approach to represent product variability. It combines the modular representation of changes between system variants with expressive means to capture the influence of product features. In delta modeling, a set of diverse systems is represented by a designated core system and a set of deltas describing modifications to the core system to implement further system variants. A particular product configuration is obtained by applying the changes specified in the applicable deltas to the core system. The concepts of delta modeling are language-independent.

In this paper, we apply delta modeling to represent variability in software architectures that are described by the existing architecture description language *MontiArc* [HKRR11]. MontiArc is designed to model architectures for asynchronously communicating (logically) distributed systems. In order to express variable MontiArc architectures by delta modeling, the modification operations that can be specified in deltas over architecture descriptions will be defined. An architectural delta can add, remove and modify components and alter the communication structure between these components. This reflects the variability that is induced by different product features. By applying a sequence of deltas to the MontiArc description of a designated core architecture, MontiArc descriptions for architectures of other product variants can be obtained. The resulting modeling language for architectural variability, $\Delta$-*MontiArc*, provides means to modularly specify architectural variability by defining architectural deltas.

$\Delta$-MontiArc supports the proactive and extractive development [Kru02] of software product line architectures. Hence, it can be used methodologically as an incremental variability model where components are added or replaced by refined and enhanced versions. This is in contrast to annotative approaches [ZHJ03, Gom04, CA05], where a model containing all possible variants is stripped down for a particular feature configuration. Furthermore, existing legacy applications can be transformed into a product line by specifying the architectural deltas that are required to describe the complete product space.

## 2 Architectural Modeling with MontiArc

Architecture description languages (*ADL*s) [MT00] support the modeling, design, analysis, and evolution of system architectures. In ADLs, architectures are generally described in terms of components, connectors and communication relationships. ADLs facilitate a high-level description of the system structures in a specific domain and support reasoning about structural system properties [GMW97].

The textual ADL *MontiArc* [HKRR11] is designed for modeling distributed information flow architectures in which communication is based on asynchronous messages. It is developed using the DSL framework MontiCore [GKR$^+$08] that supports the agile development of textual domain-specific languages. Following [MT00], architectural components in MontiArc are units of computation or storage defining their computational commitments via interfaces. Theses interfaces are the only interaction points of components to provide clear concepts of interaction between entities of computation [BS01].

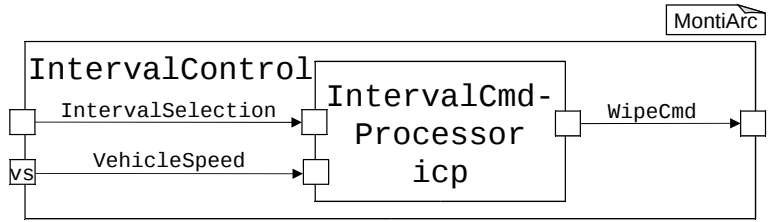

Figure 1: Architecture of the IntervalControl

As an example of a system architecture description in MontiArc, we use an excerpt of an embedded windshield wiper system. The architecture is graphically represented in Figure 1. The component `IntervalControl` receives an interval selection from the driver and emits a command `WipeCmd` to control a connected wiper actuator. It is hierarchically decomposed into a subcomponent `IntervalCmdProcessor` named `icp` that calculates the desired wiper behavior based on the selection and the vehicle speed. In MontiArc, a subcomponent is either an instance of a component referenced from a separate model, or a local component definition (similar to a private inner classes in object-oriented programming languages). The interface definition of the component `IntervalControl` contains an incoming port `IntervalSelection`, an incoming port `vs`, and an outgoing port `WipeCmd`. All ports are connected to the subcomponent `icp` using connectors. Ports and subcomponents obey an implicit naming rule. If the used type is unique in the current component definition, the usage of an explicit name is optional and the artifact is implicitly named after its type.

Listing 1 shows the textual description of the above described architecture (cf. Figure 1) in MontiArc syntax. Components are organized in packages (l. 1). A component defini-

```
package wipe;

component IntervalControl {
    autoconnect port;

    port
        in IntervalSelection,
        in VehicleSpeed vs,
        out WipeCmd;

    component IntervalCmdProcessor icp;

    connect vs -> icp.VehicleSpeed;
}
```

Listing 1: Structural component `IntervalControl` in MontiArc syntax

tion consists of the keyword `component` followed by its name and curly brackets that surround the component's architectural elements. As in the graphical notation, the component's interfaces are described by ports (ll. 6-9). The keyword `port` and the direction (in or out) are followed by a communication data type and the port's name (optional). Subcomponents that instantiate other component definitions are created with the keyword `component` followed by the subcomponent's type and an optional name (l. 11). Subcomponents that define an inner component are created with the same component definition syntax as described above.

MontiArc offers several mechanisms to connect subcomponents: the `autoconnect port` (l. 4) statement automatically connects all type-compatible ports with the same name. If the autoconnect statement is parametrized with the keyword type, ports with the same type are connected. Using this mechanism, the connector from the outgoing port `WipeCmd` of component `icp` to the outgoing port `WipeCmd` is created in the example. If it is not possible to create all connections automatically as uniquely identifying names may not be always given, explicit connections can be created using the `connect` statement (l. 13) connecting one source port with one or more target ports. If a target or source port belongs to a subcomponent, it is qualified with the subcomponent's name. Implicit connectors can always be redefined by explicit connector definitions.

## 3  Delta Modeling

Delta modeling [CHS10, SBB$^+$10, Sch10] is a transformational approach for modular variability modeling. The concepts of delta modeling are language-independent. A *delta-oriented product line* is split into a core model and a set of model deltas that are developed during domain engineering [PBvdL05]. The *core model* corresponds to a product for some valid feature configuration. The core model can be developed according to well-established single application engineering principles or derived from an existing legacy system. The variability of the product line is handled by model deltas. The *model deltas* specify modifications to the core model in order to integrate other product features. The modifications comprise additions of modeling entities, removals of modeling entities, and modifications of modeling entities by changing the internal structure of these entities. The model deltas contain *application conditions* determining under which feature configurations the specified modifications have to be carried out. These application conditions are Boolean constraints over the features in the feature model and build the connection between the feature model and the variability of the product artifacts. A model delta does not necessarily refer to exactly one feature, but potentially to a combination of features which allows handling the influence of feature combinations to the artifacts individually. The number of model deltas that have to be created to cover the complete product space depends on the desired granularity of the application conditions.

A model for a particular feature configuration is obtained by *delta application*. The modifications specified by the model deltas which have a valid application condition for the respective feature configuration are applied to the core model. To avoid conflicts between modifications targeting the same model entities, a partial order between the model deltas can be defined that determines in which order the model deltas have to be applied if applied together. The partial order captures the necessary dependencies in order to guarantee that for every feature configuration a unique model can be generated. Furthermore, the partial order ensures that a model delta is applicable to the core model or an intermediate model during delta application. *Applicability* requires that added model entities do not exist and removed and modified entities exist. An intermediate model during delta application may not be well-formed according to the well-formedness rules of the underlying modeling language. However, after all applicable model deltas are applied, the resulting product model must be well-formed.

# 4  Architectural Variability Modeling in $\Delta$-MontiArc

In order to make the ADL MontiArc presented in Section 2 more amenable to product line engineering, we apply delta modeling presented in Section 3 to MontiArc. This results in the architectural variability modeling language $\Delta$-MontiArc. The modification operations that can be specified in model deltas in $\Delta$-MontiArc allow the addition, removal and modification of all architectural elements of MontiArc, i.e., subcomponents, associated ports and connectors. The replace operation substitutes a subcomponent by another subcomponent with the same interface. This allows changing the internal realization of a subcomponent.

A MontiCore [GKR+08] grammar defining the syntax of $\Delta$-MontiArc is shown in Listing 2. Based on MontiCore's reuse mechanisms, the existing MontiArc grammar can be extended by the `MADelta` production defining model deltas for MontiArc descriptions. The production `ArcElement` for representing architectural elements, like components, ports, or connectors, in MontiArc is also reused. A model delta in $\Delta$-MontiArc starts with the keyword `delta` (l. 2) followed by the name of the delta. Using the optional `after` clause (l. 3), a partial order on the application of model deltas can be defined. The keyword `after` is followed by a list of model delta names. The defined model delta has to be applied after the listed deltas during delta application, if the deltas are applied together. The keyword `when` is followed by a logical constraint (l. 5) which defines the application condition of the model delta. The `Constraint` production represents Boolean constraints over the feature model and is omitted for space reasons. The changes of a MontiArc architecture specified by a model delta are defined using the production `MADeltaBody` (l. 7) which may contain an optional (?) `ExpandAutoConStatement` and an arbitrary number (*) of statements defining modifications of architectural elements. The `ExpandAutoConStatement` (ll. 20-21) triggers the MontiArc autoconnect mechanism after the application of delta modifications. The interface `Statement` is implemented by several productions that allow modifying, adding, or removing ArcElements from the target model (ll. 11-16). The `ReplaceStatement` (ll. 17-19) removes a subcomponent `oldComp` and replaces it with a new subcomponent `newComp`.

A $\Delta$-MontiArc product line model consists of the core model which is a standard MontiArc architecture for the core feature configuration and a set of model deltas specfied using the above syntax. A MontiArc architecture for a particular feature configuration is

```
1  MADelta =
2    "delta" Name
3    ("after" predecessors:QualifiedName
4        ("," predecessors:QualifiedName)? )?
5    "when" Constraint MADeltaBody;
6
7  MADeltaBody = "{" ExpandAutoConStatement? Statement* "}";
8
9  interface Statement;
10
11 ModifyStatement implements Statement =
12     "modify" ArcElement ";";
13 AddStatement implements Statement =
14     "add" ArcElement ";";
15 RemoveStatement implements Statement =
16     "remove" ArcElement ";";
17 ReplaceStatement implements Statement =
18     "replace" oldComp:ArcReference
19     "with" newComp:ArcReference ";";
20 ExpandAutoConStatement =
21     "expand autoconnect;";
```

Listing 2: MontiCore grammar for the $\Delta$-MontiArc language

obtained from a $\Delta$-MontiArc product line model as follows: First, determine the model deltas that have a valid application condition in the `when` clause for the given feature configuration. Second, find a linear order of the applicable deltas that is compatible with the partial order of delta application specified in the `after` clauses. Third, apply the modifications specified in the model deltas one by one in the given order to the core model. If the `expand autoconnect` statement is provided, matching ports are connected using the autoconnect mechanism, after the architectural modifications specified in the delta.

In order to ensure that the application of every model delta is defined, the following conditions must hold for each model delta during delta application. In the current stage of language definition, we use a rather restrictive approach. (1) A subcomponent named $sc$ must not be added to component $c$, if $c$ already contains a subcomponent named $sc$. (2) A port named $p$ must not be added to component $c$, if $c$ already contains a port named $p$. (3) A connector with the target port $tp$ must not be added to component $c$, if $c$ already contains a connector with the target $tp$, or if $tp$ does not exist (as a port in $c$ or a port in a subcomponent of $c$). (4) An architectural element named $ae$ must not be removed from component $c$, if $c$ does not contain an architectural element named $ae$. A port named $p$ must not be removed from component $c$, if $c$ containts a connector with $p$ as its source or target. (5) A subcomponent named $sc$ must not be removed from component $c$, if $c$ contains a connector that has a port that belongs to $sc$ as its source or target. (6) An architectural element named $ae$ must not be modified, if an architectural element named $ae$ does not exist. (7) The operator `replace` can only substitute a subcomponent $sc_1$ of component $c$ with a subcomponent $sc_2$ (specified by the `Target` construct), if $sc_2$ has the same interface like $sc_1$. The new component $sc_2$ will be connected as the component $sc_1$.

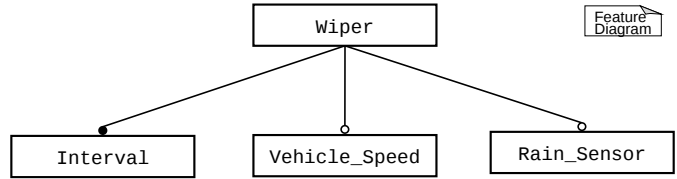

Figure 2: Feature model for the IntervalControl

```
delta DRainSensor when Rain_Sensor {

 expand autoconnect;

 modify component IntervalControl {
    add port in RainSensorStat;
    add component RainEval;
  };

  modify component IntervalCmdProcessor {
    add port in RainIntensity;
  };

}
```

Listing 3: Model delta adding the `Rain_Sensor` feature

**Example** To illustrate architectural variability modeling using ∆-MontiArc, we transform the component `IntervalControl` presented in Section 2 into a product line of similar components. The product line of similar components can be described by the feature model shown in Figure 2. All components have a mandatory feature `Interval` which allows selecting the interval for the windshield wipers. Optional features are `Vehicle_Speed` which adjusts the wiping speed based on the vehicle speed and a `Rain_Sensor` which determines the wiping based on the current rain intensity. As core model for the delta-oriented product line, we choose the component `IntervalControl` shown in Listing 1 implementing the the features `Interval` and `Vehicle_Speed`.

Listing 3 shows the model delta `DRainSensor` adding the `Rain_Sensor` feature. The model delta adds a new component `RainEval` and an incoming port `RainSensor-Stat` for the input from a rain sensor. Furthermore, an incoming port `RainIntensity` is added to the component `IntervalCmdProcessor` receiving the output from the component `RainEval`. The connections between the added components and ports are created by the `expand autoconnect` statement. Applying the delta `DRainSensor` to the core model results in a system with all three features.

In order to obtain a system which only realizes the mandatory feature, we have to provide a model delta that removes the feature `Vehicle_Speed`. This delta `DRemSpeed` is shown in Listing 4. It removes the ports `VehicleSpeed` from the sub- and supercomponents of the core model. The associated connector is also removed due to the usage of

```
1  delta DRemSpeed when !Vehicle_Speed {
2
3    expand autoconnect;
4
5    modify component IntervalCmdProcessor {
6      remove port in VehicleSpeed;
7    };
8
9    modify component IntervalControl {
10     remove port in VehicleSpeed;
11   };
12
13 }
```

Listing 4: Model delta removing `Vehicle_Speed` feature

`expand autoconnect`. In order to obtain the fourth possible system variant, realizing the feature `Rain_Sensor`, but not the feature `Vehicle_Speed`, both model deltas have to be applied the core model.

## 5 Related Work

Existing approaches to express variability in modeling languages can be classified in two main directions [VG07]: annotative and compositional approaches. Annotative approaches consider one model of all products of the product line. Variant annotations, e.g., UML stereotypes [ZHJ03, Gom04] or presence conditions [CA05], define which parts of the model have to be removed to derive a concrete product model. The orthogonal variability model (OVM) [PBvdL05] captures the variability of product line artifacts in a separate variability model where artifact dependencies take the place of annotations. In the Koala component model [vO05], the variability of a component architecture that contains all possible components is expressed by explicit components, called switches. Switches select between component variants in different system configurations playing the role of annotations. Compositional approaches associate model fragments with product features that are composed for a particular feature configuration. In [HW07, VG07, NK08], models are constructed by aspect-oriented composition. In [AJTK09], model fragments are composed by model superposition. In feature-oriented model-driven development [SBD07], a product model is composed from a base module and a sequence of feature modules. Additionally, model transformations [HMPO$^+$08, JWEG07] can represent variations of models, e.g., in [TM10, WF02], architectural variability is captured by graph transformation rules. Delta modeling [CHS10, SBB$^+$10, Sch10] can be seen as a transformational approach which provides a modular definition of transformations dependent on feature configurations.

For architectural variability modeling in [MKM06], a resemblance operator is provided that allows creating a component that is a variant of an existing component by adding, deleting, renaming or replacing component elements. The old and the new component can be used together to build further components. Instead, in delta modeling, the existing component is destroyed since a complete model of the resulting architecture will be generated by delta application. In [HvdH07], architectural variability is represented by change sets containing additions and removals of components and component connections that are applied to a base line architecture. Relationships between change sets specify which change sets may be applied together similar to application conditions of model deltas. However, the order in which change sets are applied cannot be explicitly specified. Conflicts between change sets have to be resolved by excluding the conflicting combination using a relationship and providing a new change set covering the combination which may lead to a combinatorial explosion of change sets to represent all possible variants.

# 6   Conclusion

In this paper, we have presented a first version of $\Delta$-MontiArc to support the flexible modular modeling of architectural variability by extending the existing ADL MontiArc with the concepts of delta modeling. In order to guarantee the the uniqueness of the resulting artifacts we plan to apply the criteria given in [CHS10] to $\Delta$-MontiArc. We expect to get more experience from case studies to investigate how to optimize this integration and compare it with other similar approaches. We are in particular interested in restricted forms of modifications that add and refine architectures. An appropriate refinement calculus for this is given in [PR99]. Additionally, we aim at developing a tool infrastructure for $\Delta$-MontiArc. Common language artifacts, like symbol tables, context condition checkers and model editors, as well as a generator that creates a MontiArc model for a particular feature configuration by delta application have to be developed. $\Delta$-MontiArc is a promising language for modeling architecture evolution since delta modeling is flexible enough to deal with anticipated and unanticipated variability.